\documentclass[10pt,twocolumn,prl,aps,amssymb,amsmath,tightenlines]{revtex4}

\newcommand{\half}{\mbox{$\textstyle\frac{1}{2}$}}

\begin{document}

\title{Scalar Quantum Field Theory with Cubic Interaction}

\author{Carl~M.~Bender${}^*$, Dorje~C.~Brody, and Hugh~F.~Jones}

\affiliation{Blackett Laboratory, Imperial College, London SW7 2BZ, UK}

\date{\today}

\begin{abstract}
In this paper it is shown that an $i\varphi^3$ field theory {\it is} a
physically acceptable field theory model (the spectrum is positive and the
theory is unitary). The demonstration rests on the perturbative construction of
a linear operator $\mathcal{C}$, which is needed to define the Hilbert space
inner product. The $\mathcal{C}$ operator is a new, time-independent observable
in $\mathcal{PT}$-symmetric quantum field theory.
\end{abstract}

\pacs{11.30.Er, 11.10.Lm, 12.38.Bx, 2.30.Mv}

\maketitle

A scalar $g\varphi^3$ field theory is often used as a pedagogical example of
perturbative renormalization even though this model is not physically
realistic (the energy is not bounded below). However, we argue that when $g=i
\epsilon$ is imaginary, one obtains a fully acceptable quantum field theory and
we show how to construct perturbatively the Hilbert space in which cubic scalar
field theories in $(D+1)$-dimensional Minkowski space-time are {\it
self-adjoint}. Consequently, such theories have positive spectra and exhibit
unitary time evolution. We consider here such complex field-theoretic
Hamiltonians as
\begin{eqnarray}
H=\int d{\bf x}\left[\half\pi^2_{\bf x}+\half(\nabla\varphi_{\bf x})^2
+\half\mu^2\varphi^2_{\bf x}+i\epsilon\varphi^3_{\bf x}\right],
\label{e1}
\end{eqnarray}
where $\int d{\bf x}=\int d^Dx$. [We suppress the time variable $t$ in the
fields and write $\varphi ({\bf x},t)=\varphi_{\bf x}$.] The fields in
(\ref{e1}) satisfy the ETCR $[\varphi({\bf x},t),\pi({\bf y},t)]=i\delta(
{\bf x}-{\bf y})$. As in quantum mechanics, where the operators $x$ and $p$
change sign under parity reflection $\mathcal{P}$, we assume that the fields are
{\it pseudoscalars} and also change sign under $\mathcal{P}$: $\mathcal{P}
\varphi({\bf x},t)\mathcal{P}=-\varphi(-{\bf x},t)$ and $\mathcal{P}\pi({\bf x},
t)\mathcal{P}=-\pi(-{\bf x},t)$. Field-theory models like that in (\ref{e1})
are of physical interest because they arise in Reggeon field theory and in the
study of the Lee-Yang edge singularity \cite{r-1}.

This paper is motivated by the observation that the cubic complex
quantum-mechanical Hamiltonian
\begin{eqnarray}
H=\half p^2+\half \mu^2x^2+i\epsilon x^3 \quad(\epsilon~{\rm real})
\label{e2}
\end{eqnarray}
has a positive real spectrum \cite{r0,r1,r2}. Although this Hamiltonian is not
Hermitian in the conventional sense, where Hermitian adjoint means complex
conjugate and transpose, it still defines a unitary theory of quantum mechanics
\cite{r3,r4}. This is because $H$ in (\ref{e2}) is self-adjoint with respect to
a new inner product that is distinct from the inner product of ordinary quantum
mechanics \cite{r5}.

For quantum-mechanical theories like that in (\ref{e2}) two issues need to be
addressed. First, the spectrum of $H$ must be shown to be real and positive. In
Ref.~\cite{r3} it was observed that spectral positivity of $H$ is associated
with unbroken space-time reflection symmetry ($\mathcal{PT}$ symmetry). [The
term {\it unbroken} $\mathcal{PT}$ symmetry means that every eigenstate of $H$
is also an eigenstate of ${\cal PT}$. This condition guarantees that the
eigenvalues of $H$ are real; $H$ in (\ref{e2}) has an unbroken $\mathcal{PT}$
symmetry for all real $\epsilon$.] Second, it is necessary to construct a
Hilbert space with a positive norm on which the Hamiltonian acts and to show
that $H$ is self-adjoint with respect to this nonstandard inner product. Using a
coordinate-space representation, this construction was made in Ref.~\cite{r3} by
introducing a new linear operator $\mathcal{C}$:
\begin{eqnarray}
\mathcal{C}(x,y)\equiv\sum_{n=0}^\infty\phi_n(x)\phi_n(y),
\label{e3}
\end{eqnarray}
where $\phi_n(x)$ ($n=0,1,2,\ldots$) are the eigenstates of $H$.

For the cubic Hamiltonian in (\ref{e2}), $\phi_n(x)$ satisfies the Schr\"odinger
equation
\begin{equation}
-\half\phi_n''(x)+\half\mu^2x^2\phi_n(x)+i\epsilon x^3\phi_n(x)=E_n\phi_n(x)
\label{e4}
\end{equation}
and the
boundary conditions $\phi_n(x)\to0$ as $x\to\pm\infty$. The eigenstates of $H$
are also eigenstates of $\mathcal{PT}$ and are normalized so that
$\mathcal{PT}\phi_n(x)=\phi_n^*(-x)=\phi_n(x)$. The inner product involves
$\mathcal{CPT}$ conjugation
$$\langle \psi|\chi\rangle_{\mathcal{CPT}}=\int dx\,\psi^{\mathcal{CPT}}(x)
\chi(x),$$
where $\psi^{\mathcal{CPT}}(x)=\int dy\,{\mathcal{C}}(x,y)\psi^*(-y)$, instead
of conventional Hermitian conjugation, where the inner product is $\langle\psi|
\chi\rangle=\int dx\,\psi^*(x)\chi(x)$. The novelty of these complex
Hamiltonians is that {\it the Hilbert space inner product is not prespecified;
rather, it is dynamically determined by $H$.} This new kind of quantum theory is
a sort of ``bootstrap'' theory because one must solve for the eigenstates of $H$
before knowing what the Hilbert space and the associated inner product of the
theory are.

The key problem in understanding a complex Hamiltonian like that in (\ref{e1})
or (\ref{e2}) is to determine the operator $\mathcal{C}$. In Ref.~\cite{r6}
perturbative methods were used to calculate $\mathcal{C}$ to third order in
$\epsilon$ for the $H$ in (\ref{e2}). The procedure was first to solve the
Schr\"odinger equation (\ref{e4}) for $\phi_n(x)$ as a series in powers of
$\epsilon$. Then, $\phi_n(x)$ was substituted into (\ref{e3}) and the summation
over $n$ was performed to obtain $\mathcal{C}(x,y)$ to order $\epsilon^3$. The
result was complicated, but it simplified when $\mathcal{C}$ was rewritten in
exponential form:
\begin{eqnarray}
\mathcal{C}(x,y)=\left(e^{\epsilon Q_1+\epsilon^3 Q_3+\ldots}\right)\delta(x+y)+
\mathcal{O}(\epsilon^5),
\label{e5}
\end{eqnarray}
where $Q_{2n+1}(x,p)$ are differential operators depending on $x$ and $p=-i\frac
{d}{dx}$. In the representation (\ref{e5}) only odd powers of $\epsilon$ appear
in the exponent, the coefficients are real, and the derivative operators act on
the parity operator ${\mathcal{P}}=\delta(x+y)$. Also, $\mathcal{CP}=e^{\epsilon
Q_1+\epsilon^3 Q_3+\cdots}$ is Hermitian.

Calculating $\mathcal{C}$ by direct evaluation of the sum in (\ref{e3}) is
difficult in quantum mechanics because it is necessary to determine all the
eigenfunctions of $H$. In quantum field theory such a procedure is impossible
because there is no analog of the Schr\"odinger eigenvalue problem (\ref{e4}).

The breakthrough reported here is the discovery of a new and powerful method for
calculating $\mathcal{C}$ by seeking an operator representation of it in the
form $\mathcal{C}=e^{Q(x,p)}\mathcal{P}$. This representation, which is
suggested by (\ref{e5}), has the advantage that $Q(x,p)$ is determined by
elementary operator equations so that the eigenfunctions of $H$ are not needed
to find $Q$. Thus, the technique introduced here generalizes to quantum field
theory.

We illustrate the generality of the representation $\mathcal{C}=e^Q\mathcal{P}$
by using two elementary examples:

{\it Example 1}: The complex $2\times2$ Hamiltonian
\begin{eqnarray}
H=\left(\begin{array}{cc} re^{i\theta}&s\cr s & re^{-i\theta}\end{array}\right),
\label{e6}
\end{eqnarray}
taken from Ref.~\cite{r3}, is $\mathcal{PT}$ symmetric, where ${\mathcal{P}}=
\left({0~1}\atop{1~0}\right)$ and $\mathcal{T}$ is complex conjugation. A nice
way to express $\mathcal{C}$ in the region $s^2\geq r^2\sin^2\theta$ of unbroken
$\mathcal{PT}$ symmetry is
$$\mathcal{C}=e^Q\mathcal{P},\quad Q=\half\sigma_2\ln\left[(1-\sin\alpha)/
(1+\sin\alpha)\right],$$
where $\sin\alpha=(r/s)\,\sin\theta$ and $\sigma_2=\left({0~-i}\atop{i~\,~0}
\right)$ is the Pauli matrix. As $\theta\to0$, $H$ becomes Hermitian and
$\mathcal{C}\to\mathcal{P}$.

{\it Example 2}: The Hamiltonian $H=\half p^2+\half x^2+i\epsilon x$ has an
unbroken $\mathcal{PT}$ symmetry for all real $\epsilon$. Its eigenvalues
$E_n=n+\frac{1}{2}+\frac{1}{2}\epsilon$ are all real. The $\mathcal{C}$ operator
is given exactly by $\mathcal{C}=e^Q\mathcal{P}$, where $Q=-\epsilon p$. Again,
in the limit $\epsilon\to0$ the Hamiltonian becomes Hermitian and $\mathcal{C}
\to\mathcal{P}$.

Our new procedure for constructing $\mathcal{C}$ for a
given $H$ is based on three observations made in Ref.~\cite{r3}:
(i) $\mathcal{C}$ commutes with the space-time reflection operator $\mathcal{P
T}$, but not with $\mathcal{P}$ or $\mathcal{T}$ separately; (ii) the square of
$\mathcal{C}$ is the identity; (iii) $\mathcal{C}$ commutes with $H$. To
summarize,
\begin{eqnarray}
({\rm i})~[\mathcal{C},\mathcal{PT}]=0,\quad({\rm ii})~\mathcal{C}^2={\bf 1},
\quad({\rm iii})~[\mathcal{C},H]=0.
\label{e7}
\end{eqnarray}
Substituting $\mathcal{C}=e^Q\mathcal{P}$ into condition (i) in (\ref{e7}), we
obtain $e^{Q(x,p)}=\mathcal{PT}e^{Q(x,p)}\mathcal{PT}=e^{Q(-x,p)}$, so $Q(x,
p)$ is an {\it even} function of $x$. Next, we substitute $\mathcal{C}=e^Q
\mathcal{P}$ into condition (ii) in (\ref{e7}) and get $e^{Q(x,p)}\mathcal{P}
e^{Q(x,p)}\mathcal{P}=e^{Q(x,p)}e^{Q(-x,-p)}=1$, which implies that $Q(x,p)=-Q
(-x,-p)$. Since $Q(x,p)$ is an even function of $x$, it must also be an {\it
odd} function of $p$. Finally, substituting $\mathcal{C}=e^{Q(x,p)}\mathcal{P}$
into condition (iii) in (\ref{e7}), we obtain $e^{Q(x,p)}[\mathcal{P},H]+[e^{Q
(x,p)},H]\mathcal{P}=0$. All of the Hamiltonians $H$ considered in this paper
have the form $H=H_0+\epsilon H_1$, where $H_0$ is a free Hamiltonian that
commutes with the parity operator $\mathcal{P}$ and $H_1$ represents a cubic
interaction term, which {\it anticommutes} with $\mathcal{P}$. Hence, for
quantum theories described by cubic Hamiltonians, we have
\begin{eqnarray}
\epsilon e^{Q(x,p)}H_1=[e^{Q(x,p)},H].
\label{e8}
\end{eqnarray}

For all cubic Hamiltonians, $Q(x,p)$ may be expanded as a series in {\it odd}
powers of $\epsilon$ as in (\ref{e5}):
$Q(x,p)=\epsilon Q_1(x,p)+\epsilon^3 Q_3(x,p)+\epsilon^5Q_5(x,p)+\cdots\,$.
(In quantum field theory we can interpret the coefficients $Q_{2n+1}$ as
interaction vertices of $2n+3$ factors of the quantum fields.) Substituting this
expansion into (\ref{e8}) and collecting the coefficients of like powers of
$\epsilon$, we obtain a sequence of operator equations that can be solved
systematically for the operator-valued functions $Q_n(x,p)$ $(n=1,3,5,\ldots)$
subject to the constraints that $Q(x,p)$ be even in $x$ and odd in $p$. It is
interesting that even powers of $\epsilon$ yield redundant information because
the equations arising from the coefficient of $\epsilon^{2n}$ can be derived
from the equations arising from the coefficients of $\epsilon^{2n-1},\,\epsilon^
{2n-3},\,\ldots,\, \epsilon$. The first three operator equations for $Q_n$ are
\begin{eqnarray}
\left[H_0,Q_1\right] \!&=&\! -2H_1,\nonumber\\
\left[H_0,Q_3\right] \!&=&\! -{\textstyle\frac{1}{6}}[Q_1,[Q_1,H_1]],\nonumber\\
\left[H_0,Q_5\right] \!&=&\! {\textstyle\frac{1}{360}}[Q_1,[Q_1,[Q_1,[Q_1,
H_1]]]] \nonumber\\
&&\hspace{-1cm}-{\textstyle\frac{1}{6}}\Big([Q_1,[Q_3,H_1]]
+[Q_3,[Q_1,H_1]]\Big).
\label{e9}
\end{eqnarray}

For $H$ in (\ref{e2}) we solve the equations in (\ref{e9}) by substituting the
most general polynomial form for $Q_n$ using arbitrary coefficients and then
solving for these coefficients. For example, to solve the first equation in
(\ref{e9}), $\left[H_0,Q_1\right]=-2ix^3$, we take as an {\it ansatz} for $Q_1$
the most general Hermitian cubic polynomial that is even in $x$ and odd in $p$:
$Q_1(x,p)=Mp^3+Nxpx$, where $M$ and $N$ are unknown coefficients. The operator
equation for $Q_1$ is satisfied if $M=-\frac{4}{3}\mu^{-4}$ and $N=-2\mu^{-2}$.

To present the solutions for $Q_n(x,p)$ we use $S_{m,n}$ to represent the {\it
totally symmetrized} sum over terms containing $m$ factors of $p$ and $n$
factors of $x$ \cite{r7}. Thus, $S_{0,0}=1$, $S_{0,3}=x^3$, $S_{1,1}=
\half(xp +px)$, $S_{1,2}=\frac{1}{3}\left(x^2p+xpx+px^2\right)$, and so on. We
have solved (\ref{e9}) for $Q_1$, $Q_3$, $Q_5$, and $Q_7$ in closed form in
terms of $S_{m,n}$:
\begin{eqnarray}
Q_1 \!&=&\! -{\textstyle\frac{4}{3}}\mu^{-4}p^3-2\mu^{-2}S_{1,2},\nonumber\\
Q_3 \!&=&\! {\textstyle\frac{128}{15}}\mu^{-10}p^5+{\textstyle\frac{40}{3}}
\mu^{-8}S_{3,2}+8\mu^{-6}S_{1,4}-12\mu^{-8}p,\nonumber\\
Q_5 \!&=&\! -{\textstyle\frac{320}{3}}\mu^{-16}p^7-{\textstyle\frac{544}{3}}
\mu^{-14}S_{5,2}-{\textstyle\frac{512}{3}}\mu^{-12}S_{3,4}\nonumber\\
&&\hspace{-1cm}-64\mu^{-10}S_{1,6}+{\textstyle\frac{24\,736}{45}}\mu^{-14}p^3
+ {\textstyle\frac{6\,368}{15}}\mu^{-12}S_{1,2},\nonumber\\
Q_7 \!&=&\! {\textstyle\frac{553\,984}{315}}\mu^{-22}p^9+{\textstyle\frac{97\,
792}{35}}\mu^{-20}S_{7,2}+{\textstyle\frac{377\,344}{105}}\mu^{-18}S_{5,4}
\nonumber\\
&&\hspace{-1cm}+{\textstyle\frac{721\,024}{315}}\mu^{-16}S_{3,6}+{\textstyle
\frac{1\,792}{3}}\mu^{-14}S_{1,8}-{\textstyle\frac{2\,209\,024}{105}}
\mu^{-20}p^5 \nonumber\\
&&\hspace{-1cm}-{\textstyle\frac{2\,875\,648}{105}}\mu^{-18}S_{3,2}-{\textstyle
\frac{390\,336}{35}}\mu^{-16}S_{1,4}+{\textstyle\frac{46\,976}{5}}\mu^{-18}p.
\label{e10}
\end{eqnarray}
These results constitute a seventh-order perturbative expansion of $\mathcal{C}$
in terms of the operators $x$ and $p$.

The representation $\mathcal{C}=e^Q\mathcal{P}$ bears a strong resemblance to
the WKB {\it ansatz}, which is also an exponential of a power series. Like our
calculation of $\mathcal{C}$ here, WKB methods determine the energy eigenvalues
$E_n$ {\it to all orders} in powers of $\hbar$ via a system of equations like
those in (\ref{e9}) without ever using the eigenfunctions $\phi_n(x)$ \cite{r8}.
Moreover, only the {\it even} terms in the WKB series are needed to find $E_n$;
the odd terms in the series are redundant and provide no information about
$E_n$~\cite{r8}. The difference between a conventional WKB series and the series
representation for $Q(x,p)$ is that the first term in a WKB series is
proportional to $\hbar^{-1}$ while the expansion for $Q$ contains only positive
powers of $\epsilon$. However, based on the nonperturbative calculations in
Ref.~\cite{r6}, we believe that for a $\mathcal{PT}$-symmetric $-\epsilon x^4$
theory, the first term in the expansion of $Q$ will also be proportional to
$\epsilon^{-1}$.

The massless (strong-coupling) limit $\mu\to0$ of $\mathcal{C}$ for $H$ in
(\ref{e1}) is interesting because it is singular. [The dimensionless
perturbation parameter is $\epsilon\mu^{-5/2}$, so negative powers of $\mu$
appear in every order in (\ref{e10}). Thus, as $\mu\to0$, the perturbation
series for $Q$ ceases to exist.] As $\mu\to0$, a new {\it nonpolynomial}
representation for $Q$ emerges. To find $\mathcal{C}$ when $\mu=0$ we return to
the operator equations in (\ref{e9}) and seek new solutions for the special
case $H_0=\half p^2$. The situation here is like that in Ref.~\cite{r7}, where
the objective was to calculate the time operator in quantum mechanics. Here, as
in the case of the time operator, we use Weyl ordering to generalize $S_{m,n}$
from positive $m$ to negative $m$. For example, $S_{-1,1}=\half\big(x\frac{1}
{p}+\frac{1}{p}x\big)$. The solution to the first equation in (\ref{e9}),
$\big[\half p^2,Q_1\big]=-2ix^3$, is $Q_1=\half S_{-1,4}+\alpha S_{-5,0}$,
where $\alpha$ is arbitrary. This solution is odd in $p$, even in $x$, and has
the same dimensions as $Q_1$ in (\ref{e10}). Also,
$Q_3=\left(\frac{3}{32}\!+\!\frac{305}{8}\alpha\right)S_{-11,4}-\left(\frac{135}
{16}\!+\!\frac{5773}{8}\alpha\!-\!\frac{75}{12}\alpha^2\right)S_{-13,2}
+\left(\frac{7}{16}\!+\!20\alpha\right)S_{-9,6}-\frac{3}{32}S_{-7,8}+\frac{1}{40
}S_{-5,10}+\beta S_{-15,0}$,
where $\beta$ is a new arbitrary number.

The new operator techniques introduced in this paper extend to systems having
several dynamical degrees of freedom. We have calculated $\mathcal{C}$ to third
order in $\epsilon$ for
\begin{eqnarray}
H=\half\left(p_x^2+p_y^2\right)+\half\left(x^2+y^2\right)+i\epsilon x^2y,
\label{e11}
\end{eqnarray}
which has two degrees of freedom. The result is
\begin{eqnarray}
\hspace{-2cm}Q_1(x,y,p,q)\!&=&\!-{\textstyle\frac{4}{3}}p^2q-{\textstyle\frac{1}
{3}}S_{1,1}y-{\textstyle\frac{2}{3}}x^2q,\nonumber\\
Q_3(x,y,p,q)\!&=&\!{\textstyle\frac{512}{405}} p^2q^3+{\textstyle\frac{512}
{405}} p^4q+{\textstyle\frac{1088}{405}} S_{1,1}T_{2,1} \nonumber\\
&& \hspace{-3.1cm} -{\textstyle\frac{256}{405}} p^2T_{1,2}
+ {\textstyle\frac{512}{405}} S_{3,1}y +{\textstyle\frac{288}{405}} S_{2,2}q
-{\textstyle\frac{32}{405}}x^2q^3  \nonumber\\
&&\hspace{-3.1cm} +{\textstyle\frac{736}{405}} x^2T_{1,2}
-{\textstyle\frac{256}{405}} S_{1,1}y^3 +{\textstyle\frac{608}{405}} S_{1,3}y
-{\textstyle\frac{128}{405}} x^4q -{\textstyle\frac{8}{9}}q,
\label{e12}
\end{eqnarray}
where $T_{m,n}$ represents a totally symmetric product of $m$ factors of $q$ and
$n$ factors of $y$. For the Hamiltonian
\begin{eqnarray}
H=\half\left(p_x^2+p_y^2+p_z^2\right)+\half\left(x^2+y^2+z^2\right)+i\epsilon
xyz,
\label{e13}
\end{eqnarray}
which has three degrees of freedom, we have
\begin{eqnarray}
Q_1(x,y,z,p,q,r)\!&=&\!-{\textstyle\frac{2}{3}}(yzp+xzq+xyr)
-{\textstyle\frac{4}{3}}pqr,\nonumber\\
Q_3(x,y,z,p,q,r)\!&=&\!{\textstyle\frac{128}{405}}\left(p^3qr+q^3pr+r^3qp\right)
\nonumber\\
&&\hspace{-3cm}+{\textstyle\frac{136}{405}}[pxp(yr+zq)+qyq(xr+zp)+rzr(xq+yp)]
\nonumber\\
&&\hspace{-3cm}-{\textstyle\frac{64}{405}} (xpxqr+yqypr+zrzpq)+
{\textstyle\frac{184}{405}} (xpxyz+yqyxz \nonumber\\
&&\hspace{-3cm}+zrzxy)-{\textstyle\frac{32}{405}}\big[x^3(yr+zq)+y^3(xr+zp)
\nonumber\\
&&\hspace{-3cm}+z^3(xq+yp)\big]
-{\textstyle\frac{8}{405}}\left(p^3yz+q^3xz+r^3xy\right).
\label{e14}
\end{eqnarray}

Note that the technique used in Ref.~\cite{r9} to calculate $\mathcal{C}$ gives
$Q_1$ in (\ref{e12}) and (\ref{e14}), but it becomes hopelessly difficult beyond
first order because one encounters the problems associated with degenerate
energy levels. Our new method works to all orders of perturbation theory because
the eigenfunctions are not needed to obtain $\mathcal{C}$.

We now show how to apply the powerful techniques we have developed in quantum
mechanics to the calculation of $\mathcal{C}$ in quantum field theory. As
before, we express $\mathcal{C}$ in the form $\mathcal{C}=e^{\epsilon Q_1+
\epsilon^3Q_3+\ldots}\mathcal{P}$, where now $Q_{2n+1}$ ($n=0,\,1,\,2,\,\ldots$)
are real functionals of the fields $\varphi_{\bf x}$ and $\pi_{\bf x}$. To find
$Q_n$ for $H$ in (\ref{e1}) we must again solve the operator equations in
(\ref{e9}). In terms of the inverse Green's function $G_{{\bf x}{\bf y}}^{-1}
\equiv(\mu^2-\nabla_{\!\bf x}^2)\delta({\bf x}-{\bf y})$ the first equation in
(\ref{e9}) is
\begin{eqnarray}
\left[\int\!\!d{\bf x}\,\pi^2_{\bf x}+\int\!\!\!\!\int\!\!
d{\bf x}\,d{\bf y}\,\varphi_{\bf x}G_{\bf xy}^{-1}\varphi_{\bf y},Q_1\right]=
-4i\!\!\int\!\!d{\bf x}\,\varphi^3_{\bf x}.
\label{e15}
\end{eqnarray}

Recalling the result for $Q_1$ in (\ref{e10}) we try the {\it ansatz}
\begin{eqnarray*}
Q_1=\int\!\!\!\!\int\!\!\!\!\int\!\!d{\bf x}\,d{\bf y}\,d{\bf z}\left(
M_{({\bf xyz})}\pi_{\bf x}\pi_{\bf y}\pi_{\bf z}
+N_{{\bf x}({\bf yz})}\varphi_{\bf y}\pi_{\bf x} \varphi_{\bf z}\right)\!.
\end{eqnarray*}
The notation $M_{({\bf x}{\bf y}{\bf z})}$ indicates that this function is
totally symmetric in its three arguments and the notation $N_{{\bf x}({\bf y}
{\bf z})}$ indicates that this function is symmetric under the interchange of
the second and third arguments. The unknown functions $M$ and $N$ are form
factors; they describe the spatial distribution of three-point interactions of
the field variables in $Q_1$. {\it The nonlocal spatial interaction of the
fields is an intrinsic property of $\mathcal{C}$.}

To determine $M$ and $N$ we substitute $Q_1$ into (\ref{e15}) and obtain a
coupled system of differential equations:
\begin{eqnarray}
&&\hspace{-3.2cm}
(\mu^2-\nabla_{\!\bf x}^2)N_{{\bf x}({\bf y}{\bf z})}
+(\mu^2-\nabla_{\!\bf y}^2)N_{{\bf y}({\bf x}{\bf z})}
+(\mu^2-\nabla_{\!\bf z}^2)N_{{\bf z}({\bf x}{\bf y})}\nonumber\\
\!&=&\!-6\delta({\bf x}\!-\!{\bf y})\delta({\bf x}\!-\!{\bf z}),\nonumber\\
N_{{\bf x}({\bf y}{\bf z})}+N_{{\bf y}({\bf x}{\bf z})}\!&=&
\!3(\mu^2-\nabla_{\!\bf z}^2) M_{({\bf w}{\bf x}{\bf y})}.
\label{e17}
\end{eqnarray}

We solve these differential equations by Fourier transforming to momentum space,
to obtain
\begin{eqnarray}
M_{({\bf xyz})}=-\frac{4}{(2\pi)^{2D}}\!\int\!\!\!\!\int\!\! d{\bf p}\,
d{\bf q}\frac{e^{i({\bf x}-{\bf y})\cdot{\bf p}+i({\bf x}-{\bf z})\cdot{\bf q}}}
{D({\bf p},{\bf q})},
\label{e18}
\end{eqnarray}
where $D({\bf p},{\bf q})=4[{\bf p}^2{\bf q}^2-({\bf p}\cdot{\bf q})^2]
+4\mu^2({\bf p}^2+{\bf p}\cdot{\bf q}+{\bf q}^2)+3\mu^4$ is positive, and
\begin{eqnarray}
N_{{\bf x}({\bf yz})}\!&=&\!3\left(\nabla_{\!\bf
y}\cdot\nabla_{\!\bf z}+\half\mu^2\right)M_{({\bf xyz})}.
\label{e19}
\end{eqnarray}
For the special case of a $(1+1)$-dimensional quantum field theory
the integral in (\ref{e18}) evaluates to
\begin{eqnarray}
M_{({\bf x}{\bf y}{\bf z})}=-\big(\sqrt{3}\pi\mu^2\big)^{-1}{\rm K}_0(\mu R),
\label{e20}
\end{eqnarray}
where ${\rm K}_0$ is the associated Bessel function and $R^2=\half[({\bf x}-{\bf
y})^2+({\bf y}-{\bf z})^2+({\bf z}-{\bf x})^2]$.

We can perform the same calculations for cubic quantum field theories having
several interacting scalar fields. Consider first the case of {\it two} scalar
fields $\varphi_{\bf x}^{(1)}$ and $\varphi_{\bf x}^{(2)}$ whose interaction is
governed by $H=H_0^{(1)}+H_0^{(2)}+i\epsilon\int d{\bf x}\,\big[\varphi_{\bf
x}^{(1)}\big]^2\varphi_{\bf x}^{(2)}$, which is the analog of the
quantum-mechanical theory described by $H$ in (\ref{e11}). Here,
$$H_0^{(j)}=\half\!\int\!\!d{\bf x}\, \big[\pi_{\bf x}^{(j)}\big]^2
+\half\!\int\!\!\!\!\int\!\!d{\bf x}\,d{\bf y}\, \big[G_{\bf xy}^{(j)}
\big]^{-1} \varphi_{\bf x}^{(j)}\varphi_{\bf y}^{(j)}.$$
To determine $\mathcal{C}$ to order $\epsilon$ we introduce the {\it ansatz}
\begin{eqnarray*}
Q_1\!&=&\!\!\int\!\!\!\!\int\!\!\!\!\int\!\! d{\bf x}\,d{\bf
y}\,d{\bf z}\Big[ N_{{\bf xyz}}^{(1)}\left(\pi_{\bf
z}^{(1)}\varphi_{\bf y}^{(1)}
+\varphi_{\bf y}^{(1)}\pi_{\bf z}^{(1)}\right)\varphi_{\bf x}^{(2)}\\
&&\hspace{-0.5cm}+N_{{\bf x}({\bf yz})}^{(2)}\pi_{\bf x}^{(2)}\varphi_{\bf y}^{
(1)}\varphi_{\bf z}^{(1)}+M_{{\bf x}({\bf yz})}\pi_{\bf x}^{(2)}\pi_{\bf y}^{
(1)}\pi_{\bf z}^{(1)}\Big],
\end{eqnarray*}
where $M_{{\bf x}({\bf y}{\bf z})}$, $N_{{\bf xyz}}^{(1)}$, and
$N_{{\bf x}({\bf y}{\bf z})}^{(2)}$ are unknown functions and the parentheses
indicate symmetrization. We get
\begin{eqnarray}
M_{{\bf x}({\bf yz})}=-\frac{4}{(2\pi)^{2D}}\!\int\!\!\!\!\int\!\!d{\bf q}\,
d{\bf r}\,\frac{e^{i({\bf x}-{\bf y})\cdot{\bf q}+i({\bf x}-{\bf z})\cdot{\bf r}
}}{{\mathcal{D}}({\bf q},{\bf r})},
\label{e103}
\end{eqnarray}
where ${\mathcal{D}}({\bf q},{\bf r})=4[{\bf q}^2{\bf r}^2-({\bf q}\cdot{\bf r})
^2]+4\mu_1^2({\bf q}+{\bf r})^2-4\mu_2^2{\bf q}\cdot{\bf r}-\mu_2^4+4\mu_1^2
\mu_2^2$, and
\begin{eqnarray}
N_{\bf xyz}^{(1)}&=&\left[-\nabla_{\!\bf y}^2-\nabla_{\!\bf y}\cdot\nabla_{\!\bf
z}+\half\mu_2^2\right]M_{{\bf x}({\bf yz})},\nonumber\\
N_{{\bf x}(\bf{yz})}^{(2)}&=&\left[\nabla_{\!\bf y}\cdot\nabla_{\!\bf z}-
\mu_1^2-\half\mu_2^2\right]M_{{\bf x}({\bf yz})}.
\label{e104}
\end{eqnarray}

For {\it three} interacting scalar fields whose dynamics is described by $H=H_0^
{(1)}+H_0^{(2)}+H_0^{(3)}+i\epsilon\int d{\bf x}\,\varphi_{\bf x}^{(1)}\varphi_{
\bf x}^{(2)}\varphi_{\bf x}^{(3)}$, which is the analog of $H$ in (\ref{e13}),
we make the {\it ansatz}
\begin{eqnarray*}
Q_1\!&=&\!\!\int\!\!\!\!\int\!\!\!\!\int\!\!d{\bf x}\,d{\bf y}\,d{\bf z}\Big[
N_{{\bf x}{\bf y}{\bf z}}^{(1)}\pi_{\bf x}^{(1)}\varphi_{\bf y}^{(2)}
\varphi_{\bf z}^{(3)}+N_{{\bf x}{\bf y}{\bf z}}^{(2)}\pi_{\bf x}^{(2)}
\varphi_{\bf y}^{(3)}\varphi_{\bf z}^{(1)}\\
&&\quad +N_{{\bf x}{\bf y}{\bf z}}^{(3)}\pi_{\bf x}^{(3)}
\varphi_{\bf y}^{(1)}\varphi_{\bf z}^{(2)}+M_{{\bf x}{\bf y}{\bf z}}
\pi_{\bf x}^{(1)}\pi_{\bf y}^{(2)}\pi_{\bf z}^{(3)}\Big].
\end{eqnarray*}

The solutions for the unknown functions are as follows: $M_{\bf xyz}$ is given
by the integral (\ref{e18}) with the more general formula $D({\bf p},{\bf q})=
4[{\bf p}^2{\bf q}^2-({\bf p}\cdot{\bf q})^2]+4[\mu_1^2({\bf q}^2+{\bf p}\cdot{
\bf q})+\mu_2^2({\bf p}^2+{\bf p}\cdot{\bf q})-\mu_3^2{\bf p}\cdot{\bf q}]+\mu^4
$ with $\mu^4=2\mu_1^2\mu_2^2+2\mu_1^2\mu_3^2+2\mu_2^2\mu_3^2-\mu_1^4-\mu_2^4-
\mu_3^4$. The $N$ coefficients are expressed as derivatives acting on $M$:
\begin{eqnarray*}
N_{\bf xyz}^{(1)}\!&=&\! \left[4\nabla_{\!\bf y}\cdot\nabla_{\!\bf z}+2(\mu_2^2
+\mu_3^2-\mu_1^2)\right] M_{\bf xyz}, \\
N_{\bf xyz}^{(2)}\!&=&\! \left[-4\nabla_{\!\bf y}\cdot\nabla_{\!\bf z}
-4\nabla_{\!\bf z}^2+2(\mu_1^2+\mu_3^2-\mu_2^2)\right] M_{\bf xyz}, \\
N_{\bf xyz}^{(3)}\!&=&\! \left[-4\nabla_{\!\bf y}\cdot\nabla_{\!\bf z}
-4\nabla_{\!\bf y}^2+2(\mu_1^2+\mu_2^2-\mu_3^2)\right] M_{\bf xyz}.
\end{eqnarray*}

Our new perturbative calculation of $\mathcal{C}$ for cubic quantum field
theories is an important step in our ongoing program to obtain new physical
models by extending quantum mechanics and quantum field theory into the complex
domain. The operator $\mathcal{C}$ is a new conserved quantity in quantum field
theory and this operator is required to construct observables and to evaluate
matrix elements of field operators.

We hope to generalize the breakthrough reported in this paper to noncubic
$\mathcal{PT}$-symmetric quantum field theories such as a $(3+1)$-dimensional
$-g\varphi^4$ field theory. This remarkable model field theory has a positive
spectrum, is renormalizable, is asymptotically free~\cite{r10}, and has a
nonzero one-point Green's function $G_1=\langle\varphi\rangle$. Consequently,
this model may ultimately help to elucidate the dynamics of the Higgs sector of
the standard model.

\vskip1pc CMB thanks the John Simon Guggenheim Foundation, the U.K. EPSRC, and
the U.S.~Department of Energy and DCB thanks the The Royal Society for support.

\noindent ${}^*$Permanent address: Department of Physics, Washington University,
St. Louis, MO 63130, USA.

\end{document}